\documentclass[aps,prl,showpacs,twocolumn,superscriptaddress]{revtex4-1}

\usepackage{graphicx,upgreek,amsmath,amsfonts,textcomp}

\newcommand{\fref}[1]{Fig.~\ref{#1}}

\begin{document}
\title{Multiscaling analysis of ferroelectric domain wall roughness}
\author{J. Guyonnet}
\affiliation{DPMC-MaNEP, University of Geneva, 24 Quai Ernest Ansermet, 1211 Geneva 4, Switzerland}
\author{E. Agoritsas}
\affiliation{DPMC-MaNEP, University of Geneva, 24 Quai Ernest Ansermet, 1211 Geneva 4, Switzerland}
\author{S. Bustingorry}
\affiliation{CONICET, Centro At\'omico Bariloche, 8400 San Carlos de Bariloche, R\'{\i}o Negro, Argentina}
\author{T. Giamarchi}
\affiliation{DPMC-MaNEP, University of Geneva, 24 Quai Ernest Ansermet, 1211 Geneva 4, Switzerland}
\author{P. Paruch}
\affiliation{DPMC-MaNEP, University of Geneva, 24 Quai Ernest Ansermet, 1211 Geneva 4, Switzerland}
\date{\today}
\begin{abstract}
Using multiscaling analysis, we compare the characteristic roughening of ferroelectric domain walls in Pb(Zr$_{0.2}$Ti$_{0.8}$)O$_3$ thin films with numerical simulations of weakly pinned one-dimensional interfaces. Although at length scales up to $L_{\mathrm{MA}} \geq 5$ $\upmu$m the ferroelectric domain walls behave similarly to the numerical interfaces, showing a simple mono-affine scaling (with a well-defined roughness exponent $\zeta$), we demonstrate more complex scaling at higher length scales, making the walls globally multi-affine (varying $\zeta$ at different observation length scales). The dominant contributions to this multi-affine scaling appear to be very localized variations in the disorder potential, possibly related to dislocation defects present in the substrate.
\end{abstract}
\pacs{68.35.Ct, 77.80.Dj, 68.35.Dv, 68.37.-d}
\maketitle

Domain walls separating regions with different orientations of the order parameter in ferroic materials can be seen as elastic interfaces pinned by disorder, part of a general framework applicable to systems as diverse as growth surfaces and fractures \cite{maloy_prl_92_crack}, flame and wetting fronts \cite{moulinet_distribution_width_contact_line}, and domain wall networks in the early universe \cite{garagounis_prd_03_universe_DWs}. In these systems, interfacial roughening with very different scaling properties \cite{agoritsas_physb_12_DES} is predicted as a function of their (non)equilibrium state \cite{kolton_prl_05_flat_interface} and the type of disorder \cite{nattermann_prb_90_creep_domainwall,halpin-healy_pra_91_directed_polymers,barabasi_pra_92_multifractiality}. From a practical viewpoint, understanding the behavior of the domain walls as elastic disordered systems allows a more accurate description of domain switching, growth, and stability, all key parameters for the successful implementation of (multi)ferroic devices based on domains \cite{scott_memories,bibes_natmat_08_magnetoelectric_NV} or domain walls \cite{bea_natmat_09_domainwalls_NV, catalan_rmp_12_DW_review}. More broadly, ferroic epitaxial thin films provide an excellent model system for testing theoretical predictions, since parameters such as field, temperature, and defect density can be controlled over a wide range.

Previous experimental roughening studies considered ferroic domain walls as equilibrated mono-affine interfaces collectively pinned by weak, randomly distributed disorder \cite{lemerle_prl_98_FMDW_creep,paruch_prl_05_dw_roughness_FE,paruch_dw_review_07}, for which particularly simple Gaussian scaling is expected. Whereas mono-affine interfaces present scaling properties (self-similarity) which can be described by a single scaling exponent at all length scales, in multi-affine systems, the scaling properties vary with the observation length scale, leading to a hierarchy of local scaling exponents \cite{barabasi_surface_growth_95}. Distinguishing between both cases through a multiscaling approach has proven useful in fracture studies, where multi-affine behavior was initially reported \cite{bouchbinder_prl_06_fractures_multiscaling,alava_jstatmech_06_2d_fractures}, but subsequently attributed to finite size artifacts \cite{santucci_pre_07_fracture_statistics}. However, recent atomic force microscopy (AFM) measurements have shown that the disorder potential landscape in ferroelectric thin films can in fact be very complex, with strong individual pinning centers \cite{kalinin_prl_08_single_defect,rodriguez_apl_08_DW_defect}, and local variations in the disorder strength and universality class \cite{jesse_natmat_08_SSPFM}. Potentially, these systems could therefore provide the sought-after experimental realization of more complex multi-affine scaling of interfacial roughening predicted by theory \cite{barabasi_pra_92_multifractiality}, in which a rich and diverse disorder landscape -- and possibly other interesting features -- could be accessed. The multi-affine nature of such a system could be established by multiscaling analysis \cite{santucci_pre_07_fracture_statistics} of individual domain walls or even different portions of a single domain wall, and directly compared with an ideal mono-affine model of weak collective pinning, in which the scaling behavior is exactly known.

In this paper, we report on such a study of the roughening of ferroelectric domain walls in Pb(Zr$_{0.2}$Ti$_{0.8}$)O$_3$ (PZT) thin films, compared to numerically simulated one-dimensional interfaces in random bond disorder (which preserves the local symmetry but not magnitude of the order parameter). Using multiscaling analysis, we demonstrate that the PZT domain walls can be seen as globally multi-affine interfaces composed of mono-affine segments separated by strong, highly localized variations in the disorder potential. The behavior of the mono-affine segments (with a roughness exponent $\zeta = 0.57$) is consistent with that of one-dimensional interfaces in random bond disorder up to a characteristic length scale $L_{\mathrm{MA}}\geq5$ $\upmu$m, possibly related to the strong pinning or out-of-equilibrium effects of dislocation defects.

Formally, for a roughened interface the geometrical fluctuations from an elastically optimal flat configuration at a given length scale $r$ are quantitatively described by the probability distribution function (PDF) of relative displacements $\Delta u(r)=u(z)-u(z+r)$, whose characteristic scaling properties are reflected in the behavior of its central moments \cite{agoritsas_physb_12_DES}:
\begin{equation}
S_n(r)=\overline{\left<|\Delta u(r)|^n\right>}\sim r^{n\zeta_n},
\end{equation}
where $\zeta_n$ are the associated scaling exponents for the $n$th moment, $u(z)$ the transverse displacement along the longitudinal coordinate $z$, $\langle\cdots\rangle$ an average over $z$, and $\overline{\cdots}$ an average over disorder realizations when appropriate. One-dimensional interfaces at zero temperature equilibrium in uncorrelated disorder (weak collective pinning) are well described by a Gaussian PDF \cite{halpin-healy_pra_91_directed_polymers,mezard_jdpi_91_replica,rosso_jstatmech_05_gaussian}, and are thus inherently mono-affine. In this case, the roughness function $B(r)\equiv S_2(r)_{\mathrm{mono-affine}} \sim r^{2\zeta}$ suffices to fully characterize the scaling, with a single-valued exponent $\zeta_n=\zeta$ $\forall n$. Likewise, the displacement-displacement correlation functions, equivalent to the height-height correlation functions considered in fracture surfaces,
\begin{equation}
C_n(r)=\overline{\left<|\Delta u(r)|^n\right>^{1/n}}\sim r^{\zeta_n} \sim r^{\zeta},
\end{equation}
can be collapsed to a universal curve \cite{santucci_pre_07_fracture_statistics}. The specific value of the roughness exponent $\zeta$ depends on the system dimensionality, the nature of the disorder, and the range of the elastic interactions. More complex disorder models \cite{barabasi_pra_92_multifractiality}, however, have shown multi-affine behavior characterized by an infinite set of individual scaling exponents $\zeta_n$ and a hierarchy of local roughness exponents.

\begin{figure}
\includegraphics[width=\columnwidth]{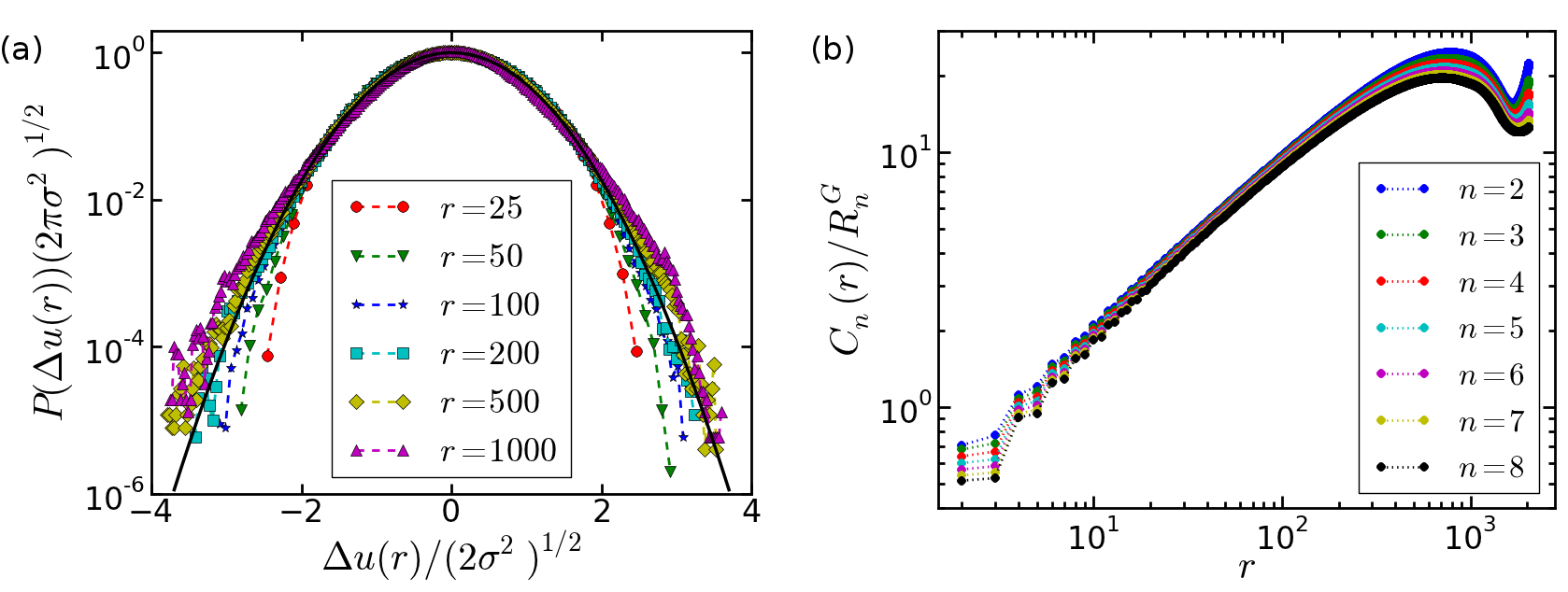}
\caption{(a) PDF of the relative displacements for different length scales taken over 10$^4$ numerical disorder configurations in Gaussian units, compared with the Gaussian function (solid line). $\sigma$ is the standard deviation. (b) Collapse of the Gaussian-normalized displacement-displacement correlation functions for orders 2 -- 8.}
\label{fig_num_ms}
\end{figure}

For the numerical study, simulations of a directed polymer model as a simple representation of a one-dimensional interface were performed on a discretized square lattice with an uncorrelated Gaussian noise distribution on each lattice site, using the solid-on-solid restriction $|u(z+1)-u(z)|=\pm1$. The equilibrium zero temperature configuration was obtained using the transfer-matrix method with a droplet geometry, i.e. with one end pinned at the origin while the other end is free. The roughness exponent $\zeta^{1D}_{RB} = 2/3$ is exactly known for this model \cite{kardar_nuclphysb_87,huse_response}. To compare its statistical properties with those of the ferroelectric domain walls, we obtained $10^4$ equilibrium configurations from independent disorder realizations for a system of size $L=2048$, and determined the $\Delta u(r)$ distributions for different length scales $r$.
As shown in \fref{fig_num_ms}a, these distributions are essentially Gaussian for intermediate length scales, with small deviations observed at very small and very large length scales. This observation is confirmed by the behavior of the displacement-displacement correlation functions, which collapse on a universal curve when renormalized by the Gaussian ratios $R_n^G=C_n^G(r)/C_2^G(r)$, independent of $r$ and $\zeta$~\cite{santucci_pre_07_fracture_statistics}. \fref{fig_num_ms}b shows such a collapse for orders $n =$ 2--8, for the numerical interfaces in the $r\in [10:500]$ intermediate regime. We attribute the slight deviations once again apparent at very small and large length scales to finite size effects. Indeed, deviations at very small length scales are a known signature of the system microstructure, in our case corresponding to the lattice discretization. At very large length scales, as $r$ approaches the system size, lack of statistics prevents sufficient averaging.

\begin{figure}[b]
\includegraphics[width=\columnwidth]{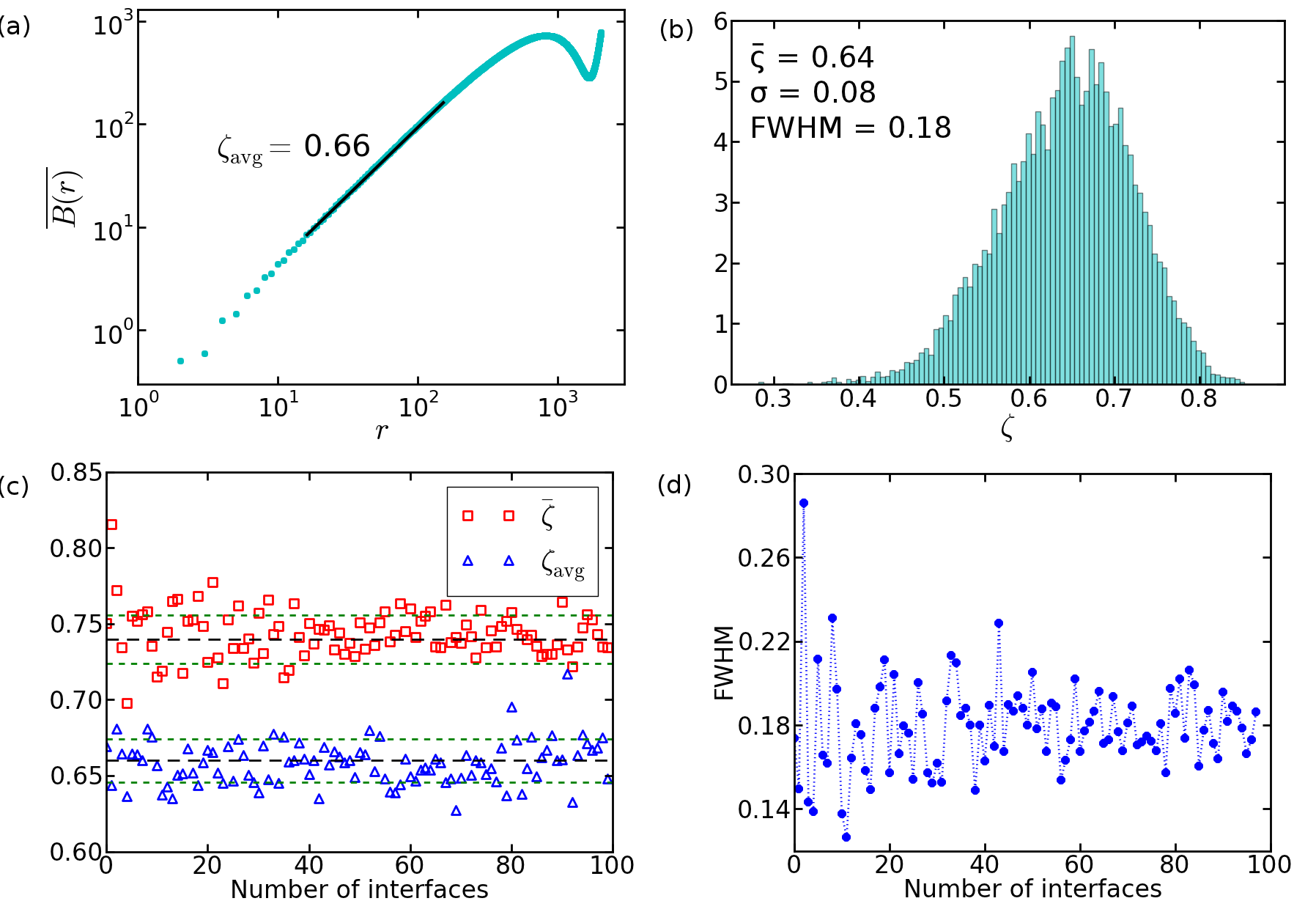}
\caption{(a) Disorder-averaged roughness function with a roughness exponent $\zeta_{\mathrm{avg}}=0.66$, and (b) roughness exponents of all individual numerical interfaces, with mean value ${\overline{\zeta}=0.64}$, both in excellent agreement with theoretical ${\zeta^{1D}_{RB}=2/3}$. (c) $\zeta_{\mathrm{avg}}$ and $\overline{\zeta}$ (shifted by 0.1 for visual clarity) for small numbers of interfaces, with the values for 10$^4$ configurations indicated by long-dashed lines, and $\pm\sigma$ by the short-dashed lines. (d) Converging FWHM of the $\zeta$ distribution for small numbers of interfaces.}
\label{fig_num_roughness}
\end{figure}

From the numerical simulations, we extract $\zeta_{\mathrm{avg}}=0.66$ from the $\overline{B(r)}$ averaged over the different disorder realizations (\fref{fig_num_roughness}a), in excellent agreement with the theoretical roughness exponent $\zeta^{1D}_{RB}=2/3$. A comparable value of $\overline{\zeta}=0.64$ is obtained from the mean of the distribution of individual roughness exponents, shown in \fref{fig_num_roughness}b. We attribute the slight negative skewness of the histogram to the local solid-on-solid restrictions, constraining the roughness exponent to $\zeta<1$, thus ``compressing'' the distribution to the right. In experimental systems presenting a symmetric $\zeta$ distribution, the direct averaging method should provide a satisfactory estimate of the characteristic roughness exponent.

An important, previously unremarked feature with significant consequences for the interpretation of both numerical and experimental results, is the large spread of the individual roughness exponent values. We emphasize here that mono-affine interfaces presenting a Gaussian PDF are expected to be fully described by a single-valued exponent. The large spread of the values extracted for $\zeta$ from individual numerical interfaces, with the distribution characterized by a full width at half maximum (FWHM) of nearly 30\% of the mean value, therefore clearly shows that there exist inherent fluctuations of the scaling behavior from one interface to the other. To obtain a meaningful value of the roughness exponent, it is therefore crucial to average over a sufficiently large data set. From the rapid convergence of the FWHM of the distribution as a function of the number of interfaces included in the average (\fref{fig_num_roughness}d), we conclude that reasonable averaging can already be obtained with a few tens of different configurations. Moreover, the fluctuations of $\zeta_{\mathrm{avg}}$ and $\overline{\zeta}$ for small numbers of interfaces are strikingly similar and equally rapidly bounded by a standard deviation $\sigma\approx0.15$ (\fref{fig_num_roughness}c).

\begin{figure*}
\includegraphics[width=\textwidth]{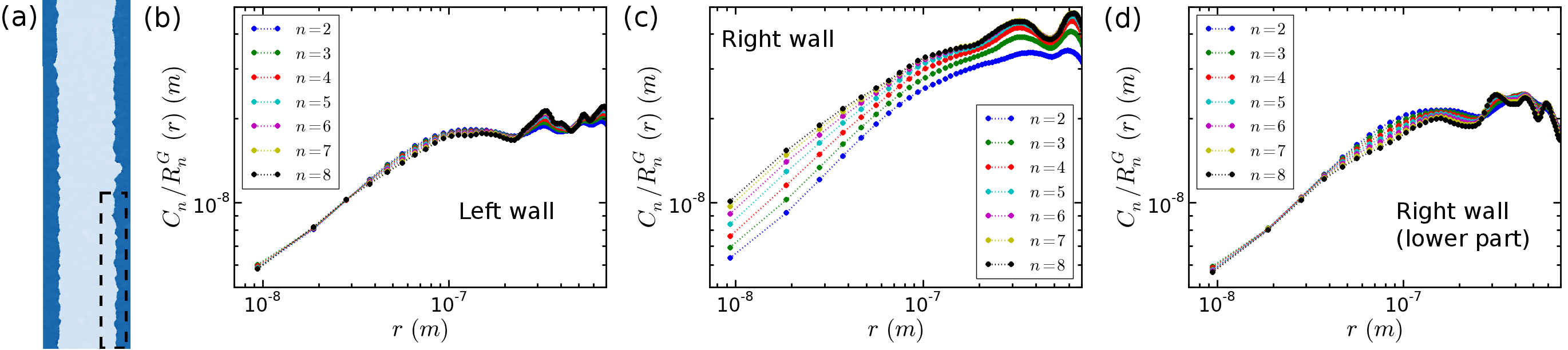}
\caption{$1.2\times4.8$ $\upmu$m$^2$ PFM phase map showing two domain walls, with light and dark contrast corresponding to up and down polarization, respectively. For the full domain walls, the normalized correlation functions show either collapse (b) or fanning (c) between different orders. Collapse is recovered (d) when considering only the lower part of the right wall (dashed box).}
\label{fig_pfm}
\end{figure*}

In parallel to these studies, experimental measurements of ferroelectric domain wall roughness were carried out on epitaxial ferroelectric PZT thin films, monodomain up-polarized as-grown, with the polarization axis perpendicular to the film plane \footnote{60--70 nm PZT films on 30--40 nm metallic SrRuO$_3$ were epitaxially grown on single-crystal (001) SrTiO$_3$ ({\protect \it CrysTec}) by off-axis radio-frequency magnetron sputtering, as detailed and characterized in \cite{gariglio_apl_07_PZT_highTc}. All the domain writing and PFM measurements were carried out in an {\protect \it Omicron VT} AFM at $5\times 10^{-10}$ mbar, using a {\protect \it Nanonis} controller and {\protect \it $\mu$Masch} DPER14 tips, at 1 V, 1 kHz AC bias, on samples introduced from ambient. Each domain wall was consecutively imaged several times, to obtain an average roughness function minimizing potential measurement artifacts such as scanner drift.}. Rectangular down-polarized domains with 180$^\circ$ domain walls were written using a positively biased scanning AFM tip, and imaged by piezoresponse force microscopy (PFM), as illustrated in \fref{fig_pfm}a, allowing the domain wall position to be precisely determined. For each domain wall, we first computed the normalized displacement-displacement correlation functions $C_n(r)/R_n^G$ for orders $n =$ 2--8. The resulting functions  show roughnening up to a length scale $r^*$ = 100 nm ($L^*$ in \cite{paruch_prb_12_quench}) and display behavior ranging from complete collapse to complete fanning, as shown for two domain walls less than 1 $\upmu$m apart in \fref{fig_pfm}a-c. The observed saturation regime above $r^*$ may be attributed to the artificial writing process, limiting domain wall roughening at higher length scales \cite{paruch_prl_05_dw_roughness_FE}. Fanning between the correlation functions of different orders may of course be induced by instrumental resolution limitations \cite{santucci_pre_07_fracture_statistics}, and in our measurements the partial fanning observed for the smallest length scales can be attributed to exactly such size effects. However, the repeated observation of radically different behaviors at intermediate length scales between simultaneously-imaged domain walls emphasizes that these features are physically meaningful: the signature of the Gaussian and non-Gaussian nature of the underlying PDF, associated with mono-affine and multi-affine interfaces, respectively.

We note, moreover, that on certain domain walls presenting fanning correlation functions when analyzed in their entirety, a collapse of the correlation functions could be recovered when only a segment of the domain wall was selected. In most cases, this selection corresponded to the exclusion of obvious local fluctuations, as demonstrated in \fref{fig_pfm}. The exclusion of a visibly larger fluctuation in the middle of the right wall promotes the global non-Gaussian signature of \fref{fig_pfm}c to the locally Gaussian signature of \fref{fig_pfm}d over the domain wall segment delimited by the short dashed box in \fref{fig_pfm}a, illustrating the strongly non-uniform and local character of the domain wall scaling properties. As expected for a multi-affine interface characterized by a varying roughness exponent $\zeta$ at different length scales, distinct values of $\zeta=$0.78, 0.72 and 0.66 were computed over segments of 1, 2 and 4 $\upmu$m centered on the middle fluctuation.

Although the microscopic origin of such multiscaling behavior remains to be elucidated, and could be related either to the presence of strong pinning centers \cite{barabasi_pra_92_multifractiality}, or to out-of-equilibrium behavior of the domain walls beyond a certain length scale \cite{paruch_prb_12_quench}, our results convincingly show that the non-Gaussian nature of the relative displacement PDF, indicated by the observed fanning of the correlation functions, corresponds to a true multi-affine state. However, smaller segments of these domain walls remain mono-affine, presenting a Gaussian PDF. In other words, there exists a characteristic length scale below which domain wall portions behave as mono-affine interfaces in weak collective pinning and above which they behave as multi-affine interfaces. We stress that, while roughening is only observed up to $ r^*$ in both cases, multi-affine behavior originates from highly localized individual fluctuations at least $L_{MA}$ apart along the length of the considered domain wall segment. In our measurements, we see at most one of the local disorder fluctuations in a typical PFM domain wall image, giving a lower bound of $L_{\mathrm{MA}}\sim5$ $\upmu$m for this length scale. One possible source for such localized fluctuations are dislocation defects propagating from the SrTiO$_3$ substrate, which can act as strong pinning sites. Transmission and scanning electron microscopy studies of selectively etched single crystal SrTiO$_3$ revealed dislocation densities of $\sim$10$^8$ cm$^{-2}$ \cite{wang_prl_98_STO_dislocations,kamaladasa_jem_11_STO_dislocations}, qualitatively in agreement with the observed disorder fluctuations. Moreover, recent in-situ transmission electron microscopy observations of ferroelectric domain switching clearly show the pinning of domain walls by individual dislocations \cite{gao_natcomm_11_TEM}. The ability to non-invasively identify these very localized features at the domain walls could be especially interesting when coupled with an investigation of their functional properties \cite{guyonnet_am_11_DW_conduction}. In particular, given that oxygen vacancies \cite{seidel_PRL_10_BFO_La, farokhipoor_12_DW_BFO} increase domain wall conduction, and dislocation cores are associated with a very high presence of oxygen vacancies \cite{jia_prl_05_STO_dislocation_core}, exploring the link between multi-affinity and domain wall current levels could be a promising research pathway.

\begin{figure}
\includegraphics[width=\columnwidth]{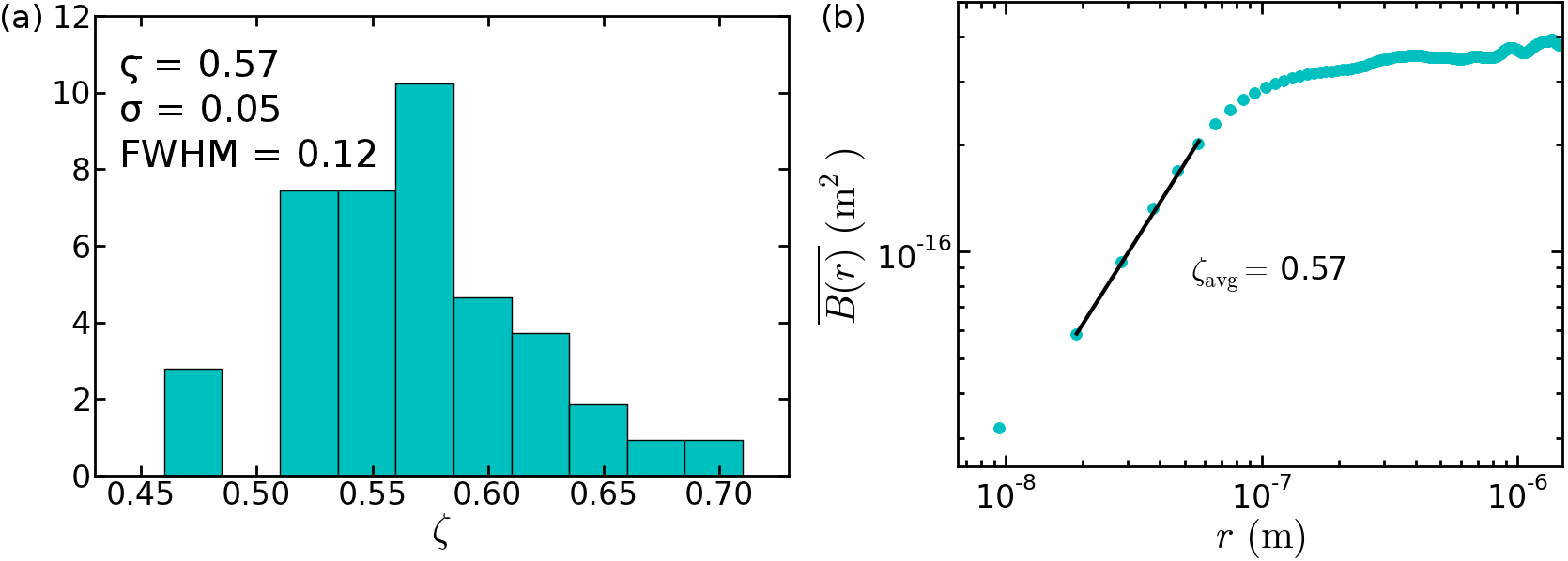}
\caption{(a) Roughness exponents obtained on a set of 43 different mono-affine domain walls with $\bar{\zeta}=0.57\pm0.05$. (b) Disorder-averaged roughness function for 22 domain walls, with similar roughness exponent $\zeta_{\mathrm{avg}}=0.57$.}
\label{fig_pfm_roughness}
\end{figure}

Finally, using multiscaling analysis, we were able to select \emph{only} mono-affine domain wall segments (on 43 out of 63 domain walls imaged), for which a single-valued roughness exponent can be defined. For these domain walls, we constructed the roughness exponent distribution, shown in \fref{fig_pfm_roughness}, with a mean value of 0.$57$ and a FWHM of 0.12, similar to the large distribution spread for the numerical interfaces, and identical to the $\zeta_{\mathrm{avg}}=0.57$, determined from $\overline{B(r)}$ (for 22 walls). 

This value is significantly higher than $\zeta \sim 0.25$ reported in previous studies of PZT domain wall roughness \cite{paruch_prl_05_dw_roughness_FE}. We believe a crucial difference between the two experiments is the use of ultra-high vacuum, eliminating the screening effects of surface adsorbates. In both cases, the saturation of the roughness function $B(r)$ clearly indicates that the artificial writing process limits domain wall roughening up to $r^*$. At ambient conditions, $r^*$ is generally smaller or comparable to the film thickness. Coupled with independent studies of their dynamics, this yields an effective domain wall dimensionality of 2.5, in good agreement with theoretical predictions for elastic two-dimensional interfaces in random bond disorder with long range dipolar interactions \cite{nattermann_jopc_83_dipole_disorder}. However, in ultra-high vacuum $r^*$ extends to further length scales, allowing the interfaces to be considered as one-dimensional, with short-range elasticity. Moreover, progressively higher temperature thermal cycling of ambient-written domain walls with initial $\zeta \sim 0.25$ also promotes roughening, increasing $\zeta$ values to $\sim 0.5$--0.6 \cite{paruch_prb_12_quench}. Although a Gaussian PDF cannot be taken to unambiguously imply that the corresponding domain walls are in fact in equilibrium, a scenario in which the mono-affine segments correspond to domain wall portions locally equilibrated with the underlying disorder landscape could agree with all the experimental observations.

The value of $\overline{\zeta} =0.57$ is close to previous reports for magnetic thin films ($\zeta_{\mathrm{avg}} = 0.69$ for 36 domain walls) \cite{lemerle_prl_98_FMDW_creep}, for shallow periodic domains in ferroelectric single crystals ($\zeta_{\mathrm{avg}} = 0.67$ for 5 domain walls) \cite{pertsev_jap_11_ceramics}, and BiFeO$_3$ thin films ($\zeta_{\mathrm{avg}} = 0.56$ for 7 domain walls) \cite{catalan_prl_08_BFO_DW}, all of which were taken to indicate one-dimensional domain walls weakly pinned by random bond disorder. However, none of these previous studies consider the possibility of deviations from mono-affine behavior, which could modify the obtained $\zeta$ values.

In conclusion, our study of the scaling properties of numerical interfaces and ferroelectric domain walls has allowed us to directly compare ideal weak collective pinning with a real system presenting strong disorder fluctuations. Using multiscaling analysis as a tool to determine the (non)Gaussian nature of the PDF of relative displacements we find that the AFM-written ferroelectric domain walls show multi-affine scaling, possibly related to the presence of dislocation defects. Between these local fluctuations smaller domain wall segments can be considered as mono-affine, with an average roughness exponent $\zeta=0.57$. 

\begin{acknowledgements}
The authors thank S. Gariglio for the PZT samples, V. Lecomte, S. Santucci and J.-M. Triscone for helpful discussions, and M. Lopes and S. Muller for technical support. UniGE work was supported by the Swiss National Science Foundation through NCCR MaNEP and Division II, and by the European Commission FP7 project OxIDes. S.B. acknowledges partial support by CONICET Grant No. PIP11220090100051.\\
Correspondence should be addressed to J.G.~(email: jill.guyonnet@unige.ch).
 \end{acknowledgements}

\end{document}